\documentclass[aps,prl,preprint,groupedaddress,showpacs,showkeys,
               preprintnumbers,floatfix]{revtex4}
\usepackage{graphicx}

\begin{document}


\title{Calculation of the $^3$He(in-flight $K^-$,$n$) reaction
for searching the deeply-bound $K^-pp$ state}


\author{Takahisa Koike}
\email[E-mail address:]{tkoike@riken.jp}
\affiliation{Advanced Meson Science Laboratory, 
RIKEN Nishina center, Wako-shi, Saitama 351-0198, Japan}
%
%
\author{Toru Harada}
\affiliation{
Research Center for Physics and Mathematics,
Osaka Electro-Communication University,
Neyagawa, Osaka, 572-8530, Japan
}

\date{\today}

\begin{abstract}
The formation of a deeply-bound $K^-pp$ state by the $^3$He(in-flight $K^-$,$n$) 
reaction is investigated theoretically in the distorted-wave impulse approximation
using the Green's function method. 
The expected inclusive and semi-exclusive spectra at $p_{K^-}$ = 1.0 GeV/c 
and $\theta_n = 0^{\circ}$ 
are calculated for the forthcoming J-PARC E15 experiment. 
We employ optical potentials between the $K^-$ and ``$pp$'' core-nucleus,
and demonstrate systematically the dependence of the spectral shape 
on $V_0$ and $W_0$, which are the real and imaginary parts of the strength 
for the optical potential, respectively. 
The necessary condition to observe a distinct peak of the $K^-pp$ 
bound state with $I=1/2$, $J^\pi=0^-$
in the spectrum turns out to be that 
the value of $V_0$ is deeper than $\sim-100$ MeV and 
$W_0$ shallower than $\sim-100$ MeV, of which 
the strength parameters 
come up to recent theoretical predictions.

\end{abstract}


\maketitle

\section{Introduction}

Over the last several years, many theoretical and experimental efforts 
have been devoted for investigations of deeply-bound kaonic 
nuclei \cite{EXA05}, 
but there is no conclusive proof on their existence 
at the present stage.
In order to study properties of antikaon $K^-/{\bar K}^0$ in 
nuclear medium, 
it is an important subject to verify the reliable evidence of deeply-bound 
kaonic nuclei because antikaons might condense in the interior 
of compact star \cite{BRL96}. 

Among various antikaon nuclear systems, 
the three-body ${\bar K}NN$ bound state with a configuration of 
$[{\bar K}\otimes \{NN \}_{I=1}]_{I=1/2}$
is expected to be the lightest and the most fundamental kaonic 
nucleus. This state is symbolically called ``$K^-pp$'' in this paper. 
In 1963, Nogami \cite{NG63} has firstly suggested 
a possible existence of the $K^-pp$ bound state by a rather crude calculation.
About 40 years later, Yamazaki and Akaishi \cite{YA02} restarted to study 
the structure of the $K^-pp$ bound state on the basis of
a quantitative few-body calculation with the Brueckner theory, 
and predicted that the binding energy and width for $K^-$ 
are $B.E. = 48$ MeV and ${\it \Gamma} = 61$ MeV, respectively.
Shevchenko, Gal and Mare$\breve{\rm s}$ \cite{She06}
performed the $\bar{K}NN$-$\pi \Sigma N$ 
coupled channel calculation for the $K^-pp$ bound state
by the Faddeev approach, and 
obtained $B.E.$ = 55-70 MeV and ${\it \Gamma}$ = 95-110 MeV.
Several theoretical works \cite{Iva05,Ike07,Dot07} 
also supported the existence of the $K^-pp$ bound state,
but the calculated binding energy and width are not converged 
theoretically.

On the other hand, FINUDA collaboration at DA$\Phi$NE \cite{Agm05}
reported experimentally evidence of a deeply-bound $K^-pp$ state 
in invariant mass spectroscopy by 
$K^-pp$ $\rightarrow$ $\Lambda$ + $p$ decay processes from 
$K^-$ absorption on $^6$Li, $^7$Li and  $^{12}$C at rest.
The measured energy and width are $B.E. = 115$ MeV 
and ${\it \Gamma} = 67$ MeV, respectively. 
However, Magas $et$ $al.$ \cite{Mag06} claimed that 
it is not necessary to postulate a $K^-pp$ bound state 
in order to explain a bump structure in the FINUDA data, 
because of the $Y$+$N$ pair interaction with
the residual nucleus, following the $K^-$  absorption 
on two nucleons in these nuclei.
Although this data analysis is continued by the FINUDA group, 
the situation is still unsettled. 
More experimental data are required in order to confirm 
whether the $K^-pp$ system has a deeply-bound state or not.

Recently, Iwasaki $et$ $al.$ \cite{Iwa06} proposed a new experiment 
searching the deeply-bound $K^-pp$ state at J-PARC, 
by the missing-mass spectrum of the $^3$He(in-flight $K^-$,$n$) 
reaction, together with the invariant-mass spectra 
detecting all particles via decay processes from the $K^-pp$ bound state
(J-PARC E15).
Our purpose is to clarify theoretically the expected inclusive 
and semi-exclusive spectra 
of the $^3$He(in-flight $K^-$,$n$) reaction for preparing the 
forthcoming J-PARC E15 experiment.

As for (in-flight $K^-$,$N$) reactions on $^{12}$C 
and $^{16}$O targets, 
Kishimoto $et$ $al$. \cite{Kis99} performed measurements
for searching deeply-bound kaonic nuclei, 
but no clear evidence was observed.
Yamagata $et$ $al.$ \cite{Yam06} calculated the 
corresponding inclusive spectra by 
$^{12}$C,$^{16}$O(in-flight $K^-$, $p$) reactions, 
using $K^-$ optical potentials based on Chiral 
unitary and phenomenological approaches; 
they suggested that any distinct peaks 
of deeply-bound kaonic states will not be observed even if the 
bound states exist. The predicted states cannot be separated 
due to their broad width.
It would be naively expected that if we choose a lighter nucleus 
as a target, a signal for formation of 
kaonic nuclei appears more clearly, 
whereas no theoretical calculation has been done for the $^3$He target
so far. 
Yamazaki and Akaishi \cite{YA06} also proposed the $K^-pp$ formation 
via $p + p \rightarrow K^+ + ``\Lambda(1405)p" \rightarrow
K^+ + K^-pp$, assuming a $\Lambda(1405)p$ doorway.
Such a $K^-pp$ experiment with a $\sim$4 GeV proton beam on 
a deuterium target is planned by FOPI collaboration 
at GSI \cite{Suz05}.

In this Letter, we demonstrate firstly the inclusive 
and semi-exclusive spectra of the $^3$He(in-flight $K^-$, $n$) reaction 
at $p_{K^-}$ = 1.0 GeV/c and $\theta_n = 0^{\circ}$ 
within the distorted wave impulse approximation (DWIA),
in order to predict the production cross section of the 
$K^-pp$ bound state and to examine 
the experimental feasibility.
As studied in Ref. \cite{Yam06}, we employ the Green's function 
method by Morimatsu and Yazaki \cite{Mor94} in the DWIA framework, 
which describes well unstable hadron nuclear systems, 
comparing theoretical spectra with experimental 
ones \cite{Tadokoro95,Harada98}.

\section{Framework}
\subsection{DWIA with the Green's function method}

In the DWIA framework \cite{Hufner74},
the inclusive double-differential cross 
section of the $^3$He(in-flight $K^-$,$n$) reaction 
at the forward direction $\theta_n=0^\circ$            
in the lab system is written \cite{Tadokoro95} as
%
\begin{eqnarray}   
\frac{d^2 \sigma}{d E_{n} d \Omega_{n}} &=&
\beta \,
\left[ \frac{d \sigma(0^\circ)}{d \Omega_{n}} \right]^{({\rm elem})} \,
S(E),
\label{eq:DWIA}
\end{eqnarray}
where 
$\left[ d \sigma(0^\circ)/d \Omega_{n} \right]^{({\rm elem})}$ is
a Fermi-averaged cross section for the elementary 
$K^- + N \to N + K^-$ forward scattering which is equivalent
to a backward $K^-N$ elastic scattering \cite{Kis99} in the lab system.
Here we consider only the non-spin-flip $K^- + n \to n + K^-$ 
reaction channel.
The kinematical factor $\beta$ \cite{Tadokoro95} expresses the translation 
from the two-body $K^-$-$N$ lab system to the $K^-$-$^3$He 
lab system \cite{Dover83}, and it is defined as 
%
\begin{eqnarray}   
\beta &=& \left\{  1 - \frac{E_n^{(2)}}{E_{K^-_{\rm res.}}^{(2)}}
\frac{p_{K^-}-p_n^{(2)}}{p^{(2)}_n} \right\}
\frac{p_n \, E_n}{p_n^{(2)} E_n^{(2)}},
\label{eq:beta}
\end{eqnarray}
where $p_{K^-}$ is the incident ${K^-}$ momentum,
and $p_n^{(2)}$ and $p_n$ ($E^{(2)}_n$ and $E_n$) denote the 
outgoing neutron momenta (energies) for a nucleon target and $^3$He 
target, respectively, in the lab frame;  
$E^{(2)}_{K^-_{\rm res.}}$ is the residual $K^-$ energy in the final state. 
$S(E)$ denotes the strength function 
of the $K^-pp$ system, which is expressed in the Green's function 
method \cite{Mor94}, as
\begin{eqnarray}   
S(E) &=& 
- \, \frac{1}{\pi} \  {\rm Im} \left[ \  \sum_{\alpha, \alpha'} 
  \int d\mbox{\boldmath{$r$}} d\mbox{\boldmath{$r'$}} 
  f^{\dagger}_{\alpha}(\mbox{\boldmath{$r$}}) 
  G_{\alpha, \alpha'}(E;\mbox{\boldmath{$r$}},\mbox{\boldmath{$r'$}})
  f_{\alpha'}(\mbox{\boldmath{$r'$}}) \ \right]
\label{eq:S(E)}
\end{eqnarray}
with 
\begin{eqnarray}   
f_{\alpha}(\mbox{\boldmath{$r$}}) &=& 
\chi^{(-)*} \left( \mbox{\boldmath{$p$}}_{n},
 \frac{M_{\rm C}}{M_{K^-pp}}\mbox{\boldmath{$r$}} \right)    
\ 
\chi^{(+)} \left( \mbox{\boldmath{$p$}}_{K^-},
 \frac{M_{\rm C}}{M_{\rm A}}\mbox{\boldmath{$r$}} \right)
\ 
\langle \alpha \, | \psi_n (\mbox{\boldmath{$r$}}) | \,  i \rangle,
\label{eq:chi+-}
\end{eqnarray}
where
$\chi^{(+)}$ ($\chi^{(-)}$) 
is distorted waves of the incoming $K^-$ (outgoing neutron) with 
the momentum $\mbox{\boldmath{$p$}}_{K^-}$ 
($\mbox{\boldmath{$p$}}_{n}$);
$M_{\rm A}$, $M_{\rm C}$ and $M_{K^-pp}$ are the masses
of the $^3$He target, the core-nucleus and the $K^-pp$ system, 
respectively.
The factor of $M_{\rm C}/M_{\rm A}$ and $M_{\rm C}/M_{K^-pp}$      
takes into account the recoil effects.
Here we assumed to use the recoil factor of 
$M_{\rm C}/M_{K^-pp}$ in not only $\chi^{(-)}$ but also $\chi^{(+)}$, 
for simplicity.
$\langle \alpha \, | \psi_n (\mbox{\boldmath{$r$}}) | \,  i \rangle$ 
is a hole-state wave function for a struck neutron in the 
target, 
where $\alpha$ denotes the complete set of eigenstates for the system.
We employ a simple (1s)$^3$ harmonic oscillator model for the $^3$He 
wave function. 
Then, the relative $2N$-$N$ wave function has the form of 
$\phi_{2N\mbox{-}N}(r) \propto \exp(-r^2/2a^2)$, 
$a = b_{N} \sqrt{3/2}$. The size parameter $b_{N}$ is taken to be 1.30 fm, 
which reproduces the recent experimental r.m.s charge radius of 
$^3$He, $\sqrt{<r^2>} = 1.94$ fm \cite{Ang04}.
In this calculation, we assumed the ``$pp$'' pair 
as a rigid core with $S_{N}=\ell_{N}=0$, for simplicity. 
This assumption would be suitable for describing the considerable 
structure of the deeply-bound $K^-pp$ state, as mentioned below. 
Therefore, Green's function for the $K^-pp$ system, $G_{\alpha, \alpha'}$,
in Eq.(\ref{eq:S(E)}) 
is obtained by solving numerically the following Klein-Gordon equation 
with a $K^-$-$pp$ optical potential $U^{\rm opt}$;
%
\begin{eqnarray}   
\{ ( E - V_{\rm Coul.}(\mbox{\boldmath{$r$}}) )^2  
 + \mbox{\boldmath{$\nabla$}}^2 - \mu^2 - 2 \, \mu \, 
   U^{\rm opt}(\mbox{\boldmath{$r$}}) \}
\, G(E;\mbox{\boldmath{$r$}},\mbox{\boldmath{$r'$}} ) 
&=& \, \delta^3(\mbox{\boldmath{$r$}}-\mbox{\boldmath{$r'$}}),
\label{eq:KG-Green}
\end{eqnarray}
where $\mu$ is the reduced mass between the $K^-$ and the ``$pp$'' 
core-nucleus, and $V_{\rm Coul.}$ the Coulomb potential 
with the finite nuclear size effect.

Moreover, the strength function $S(E)$ can be decomposed into 
two parts \cite{Mor94} as 
\begin{eqnarray}  
S(E)  = S^{\rm con}(E) + S^{\rm esc}(E),
\label{eq:Sdecomp}
\end{eqnarray}
where $S^{\rm con}(E)$ describes the $K^-$ conversion 
processes including the decay modes of $K^- pp \to \pi + \Sigma/\Lambda + N$ 
(mesonic) from the one-body absorption and 
$K^- pp \to \Sigma/\Lambda + N$ (nonmesonic) from the two-body absorption, 
and $S^{\rm esc}$(E) the $K^-$ escaping processes leaving 
the core-nucleus in the ground state: 
\begin{eqnarray}  
S^{\rm con}(E) &\equiv& - \, \frac{1}{\pi} \ 
f^{\dagger}G^{\dagger}({\rm Im} \, U^{\rm opt}) G f,
\label{eq:Scon}
\\
S^{\rm esc}(E) &\equiv& - \, \frac{1}{\pi} \ 
f^{\dagger}(1 +  G^{\dagger}U^{\rm opt \dagger})({\rm Im} \, G_0)
(1 + U^{\rm opt}G)f,
\label{eq:Sesc}
\end{eqnarray}
where $G_0$ is the free Green's function.  
Here we used an abbreviated notation for $G$, $f$ and $U^{\rm opt}$.
We evaluate the distorted waves within the eikonal 
approximation as
%
\begin{eqnarray}   
\chi^{(-)*} \left( \mbox{\boldmath{$p$}}_{n},
 \mbox{\boldmath{$r$}} \right)
&=&
\exp 
\left[ - i \mbox{\boldmath{$p$}}_{n}\cdot\mbox{\boldmath{$r$}}
- \frac{i}{v_n} 
\int^{+\infty}_{z} U_n(\mbox{\boldmath{$b$}},z')dz'
\right],
\label{eq:EikonalWave-}
\\
\chi^{(+)} \left( \mbox{\boldmath{$p$}}_{K^-},
 \mbox{\boldmath{$r$}} \right)
&=&
\exp \left[ + i \mbox{\boldmath{$p$}}_{K^-}\cdot\mbox{\boldmath{$r$}}
- \frac{i}{v_{K^-}} 
  \int^z_{-\infty} U_{K^-}(\mbox{\boldmath{$b$}},z')dz'
\right]
\label{eq:EikonalWave+}
\end{eqnarray}
with an impact parameter coordinate $\mbox{\boldmath{$b$}}$ and 
the optical potential for $\lambda=K^-$ or $n$,
\begin{eqnarray}   
U_{\lambda}(r) = -i \, {v_{\lambda} \over 2} \, 
\bar{\sigma}^{\rm tot}_{\lambda N} \, (1 - i \alpha_{\lambda N}) \, \rho(r), 
\label{eq:EikonalU}
\end{eqnarray}
where $\rho(r)$ is a nuclear density distribution, and 
$\bar{\sigma}^{\rm tot}_{\lambda N}$ and $\alpha_{\lambda N}$ 
are the isospin-averaged total cross section
and the ratio of the real to imaginary parts of the forward amplitude
for the $\lambda + N$ scattering, respectively.
These parameters are chosen to be
$\bar{\sigma}^{\rm tot}_{K^-N} =\bar{\sigma}^{\rm tot}_{nN} = 40$ mb 
and $\alpha_{K^-N}=\alpha_{nN}=0$ \cite{Kis99,Cie01}.

\subsection{Optical potential for $K^-pp$ system}

Yamazaki and Akaishi \cite{YA02} obtained the $K^-pp$ 
optical potential between the $K^-$ and the ``$pp$'' pair, 
performing a variational three-body calculation. 
This YA optical potential is
parametrized in a Gaussian form with 
a range-parameter 
$\beta = 1.09$ fm, as
%
\begin{eqnarray}
U^{\rm opt}_{K^- \mbox{-} pp} (r) 
&=& (V_0 + i \ W_0) \ \exp \left[-({r/\beta})^2\right],
\label{eq:KppOptPot}
\end{eqnarray}
where $V_0 = -300$ MeV and $W_0 = -70$ MeV.
Note that these parameters are determined within a non-relativistic framework.
Solving the Klein-Gordon equation with this potential,
we find that the $K^-$ binding energy and width account for 
$B.E. = 51$ MeV and ${\it \Gamma} = 68$ MeV, respectively.
Although these values are slightly changed from $B.E. = 48$ MeV 
and ${\it \Gamma} = 61$ MeV given in Ref. \cite{YA02}, 
we employ the YA potential having 
($V_0$, $W_0$) = ($-300$ MeV, $-70$ MeV)
as a standard one, for a first step toward 
our quantitative investigation.
Shevchenko, Gal and Mare$\breve{\rm s}$ \cite{She06} also
obtained $B.E. \sim 70$ MeV and ${\it \Gamma} \sim 110$ MeV 
as the maximum values for the $K^-pp$ bound state
by the Faddeev calculation. 
We can simulate 
$B.E. = 72$ MeV and ${\it \Gamma} = 115$ MeV as a 
complex Klein-Gordon energy,
adjusting the parameters to 
($V_0$, $W_0$) = ($-350$ MeV, $-100$ MeV)
within the potential form of Eq.(\ref{eq:KppOptPot}).
We adopt these values of $(V_0, W_0)$ as an upper limit
for the reliable parameters.

\section{Results and discussion}

Let us consider the $^3$He(in-flight $K^-$,$n$) reaction
at the incident $K^-$ momentum $p_{K^-}$=1.0 GeV/c and 
the forward direction $\theta_n = 0^\circ$,
where the lab cross section 
of the elementary reaction process is maximal \cite{Kis99,Gop77}.
The lab cross section for the $K^- + n \to n + K^-$ 
forward scattering is $\sim$27 mb/sr \cite{Gop77} in free space, 
and is reduced to $\sim$15 mb/sr 
with Fermi-averaging \cite{Rosenthal80}. 
These values are about three times larger than those 
for the $K^- + p \to p + K^-$ forward scattering for the ($K^-$, $p$) reaction 
at $p_{K^-} =$ 1.0 GeV/c, e.g., 8.8 mb/sr \cite{Yam06} in free space and 
5.2 mb/sr \cite{Cie01} with Fermi-averaging.
We adopt 15 mb/sr as a value of 
$\left[ d \sigma(0^\circ)/d \Omega_{n} \right]^{({\rm elem})}$ 
in Eq.(\ref{eq:DWIA}), 
and we notice that the momentum transfer 
$Q(0^\circ)= p_{K^-} - p_n$ is negative 
($Q(0^\circ) < 0$) which means that the residual 
$K^-$ recoils backwards relative to 
the incident $K^-$ in the lab frame, 
therefore, the kinematical factor 
$\beta$ in Eq.(\ref{eq:beta}) 
enhances the spectrum 
by a factor 1-2 depending on $p_n =$ 1.0-1.4 GeV/c.

\begin{figure}
\includegraphics[scale=0.7, angle=90]{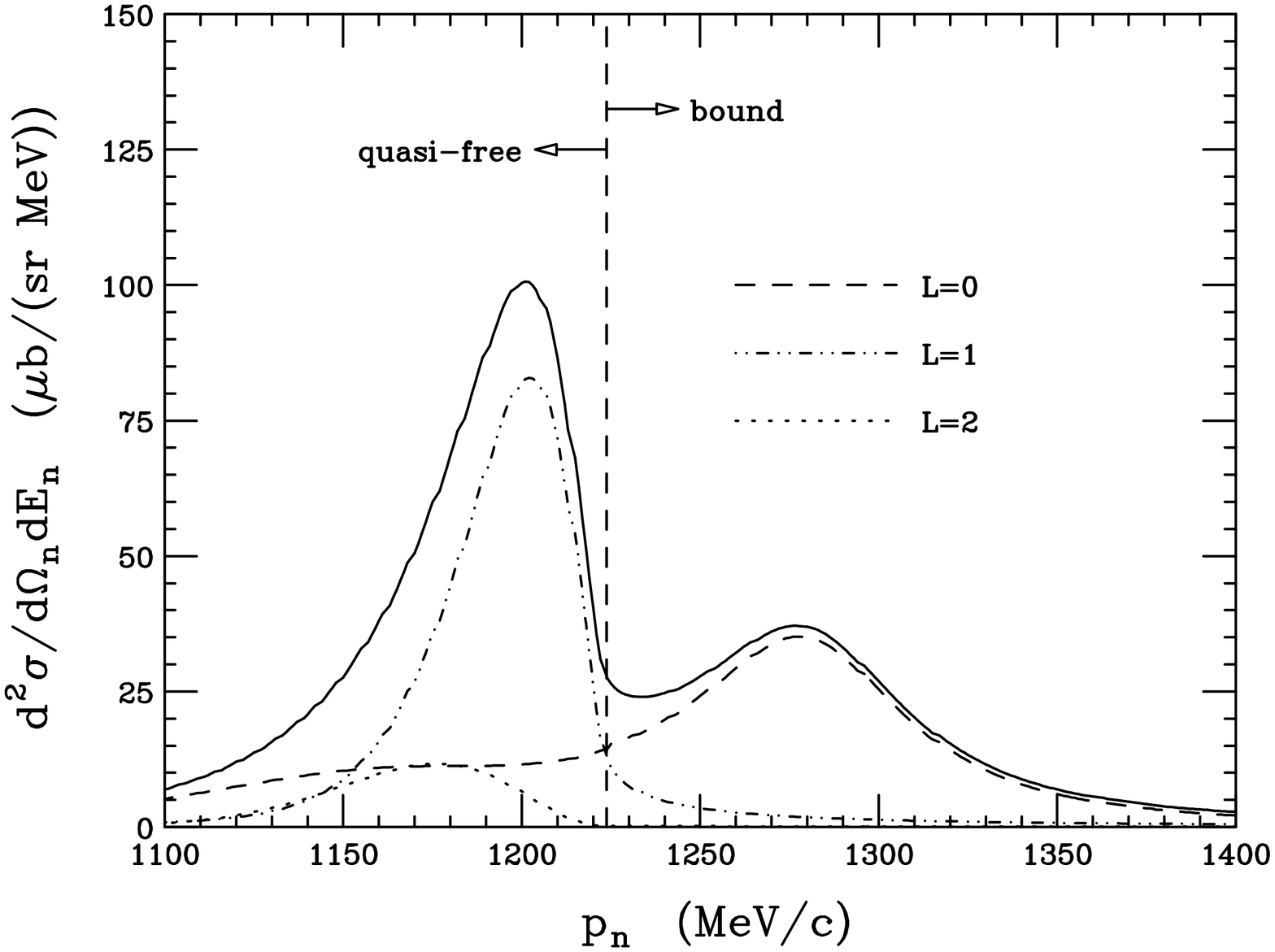}
\caption{
The calculated inclusive spectra of the $^3$He(in-flight $K^-$,$n$) 
reaction at $p_{K^-} = 1.0$ GeV/c and $\theta_n = 0^{\circ}$ 
as a function of the neutron momentum, 
using the YA optical potential with 
($V_0$, $W_0$)=($-300$ MeV, $-70$ MeV).
The vertical dashed line indicates the 
corresponding neutron momentum of $p_{n} =$ 1224 MeV/c at
the $K^-$ emitted threshold.
The contributions of partial-wave angular momentum states 
with $L$ = 0, 1 and 2 are also drawn.
\label{YAspec}}
\end{figure}

For missing-mass spectroscopy,
Fig.~\ref{YAspec} shows the calculated inclusive spectrum of 
the $^3$He(in-flight $K^-$,$n$) reaction as a function of 
the ejected neutron momentum $p_n$, and the components 
of the partial-wave angular momentum states with $L$.
Here we used the YA potential with 
($V_0$, $W_0$)=($-300$ MeV, $-70$ MeV).
We find that a distinct peak of 
the $K^-pp$ bound state with $I=1/2$, 
$J^\pi$=0$^-$ ($L = S =0$) appears around 
$p_n=$ 1280 MeV/c, which has 
$B.E. \sim$ 50 MeV and $\mit{\Gamma}\sim$ 70 MeV.
The absolute value of the integrated formation cross section 
amounts to 3.5 mb/sr. 
Note that the effective neutron number with 
the distortion for the $K^-pp$ bound state 
is $N^{\rm DW}_{1s_n \to 1s_{K^-}}$= 0.23 with $D_{\rm dis}$= 0.47, 
of which value is more than ten times as large as 
the effective proton number $P^{\rm DW}_{1p_p \to 1s_{K^-}}$= 0.013 with 
$D_{\rm dis}$=0.095 for the $K^-$ nuclear $1s$ bound state 
via the $1p_p \to 1s_{K^-}$ transition in $^{12}$C \cite{Cie01}. 
This is a major advantage of the use of the $s$-shell target nucleus 
such as $^3$He.
We stress that the calculated spectrum consists of mainly the states 
with $L = 0$ below the $K^-$ emitted threshold ($p_n =$ 1224 MeV/c), 
whereas the components of $L \geq 1$ dominates continuum states 
for the quasi-free (QF) region.
This $L=0$ dominance in the bound region is caused by a small size 
of the $s$-shell nucleus, 
contrast to the case of $p$-shell nuclei, $^{12}$C and $^{16}$O.
Moreover, the recoil effects for the light $s$-shell nuclear target 
play an important role in increasing the cross section of 
the $K^-pp$ bound state; 
if we replace the recoil factors of both $M_{\rm C}/M_{K^-pp}$ and 
$M_{\rm C}/M_{\rm A}$ by 1 in Eq.~(\ref{eq:chi+-}), 
the cross section of the peak in the bound region is reduced by half, 
whereas the yields in the QF region grow up, 
and the QF peak is shifted to the lower $p_n$ side 
and its width is broader.
When we use a reasonable factor of $M_{\rm C}/\bar{M}_{AK}$ where 
$\bar{M}_{AK}$ is the mean mass of $M_{\rm A}$ and $M_{K^-pp}$,
instead of $M_{\rm C}/M_{K^-pp}$, 
we find that the cross section around the peak in the bound region 
is enhanced by about 10\%.

--- Fig.1 ---

\begin{figure}
\includegraphics[scale=0.7, angle=90]{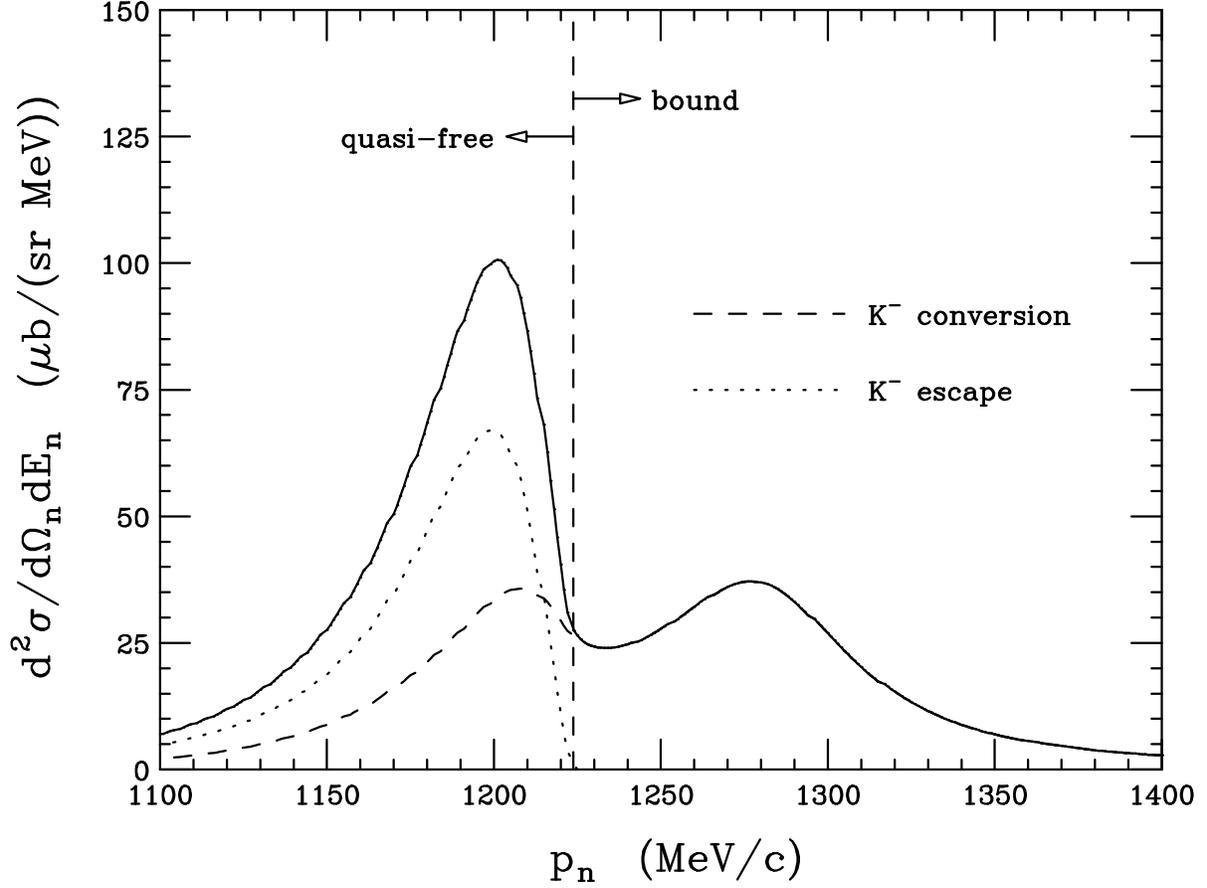}
\caption{
Same as Fig. \ref{YAspec}, but the inclusive spectrum is decomposed 
into the conversion processes of 
$K^- + ^3{\rm He} \rightarrow n + \Sigma/\Lambda + X$ (dashed line), 
and the escape processes of $K^- + ^3{\rm He} \rightarrow$ 
$n$ + $K^-$ + ``$pp$'' (dotted line). 
Note that the conversion spectrum is identical to the inclusive one 
in the bound region.
\label{YAspec2}}
\end{figure}

\begin{figure}
\includegraphics[scale=0.7, angle=90]{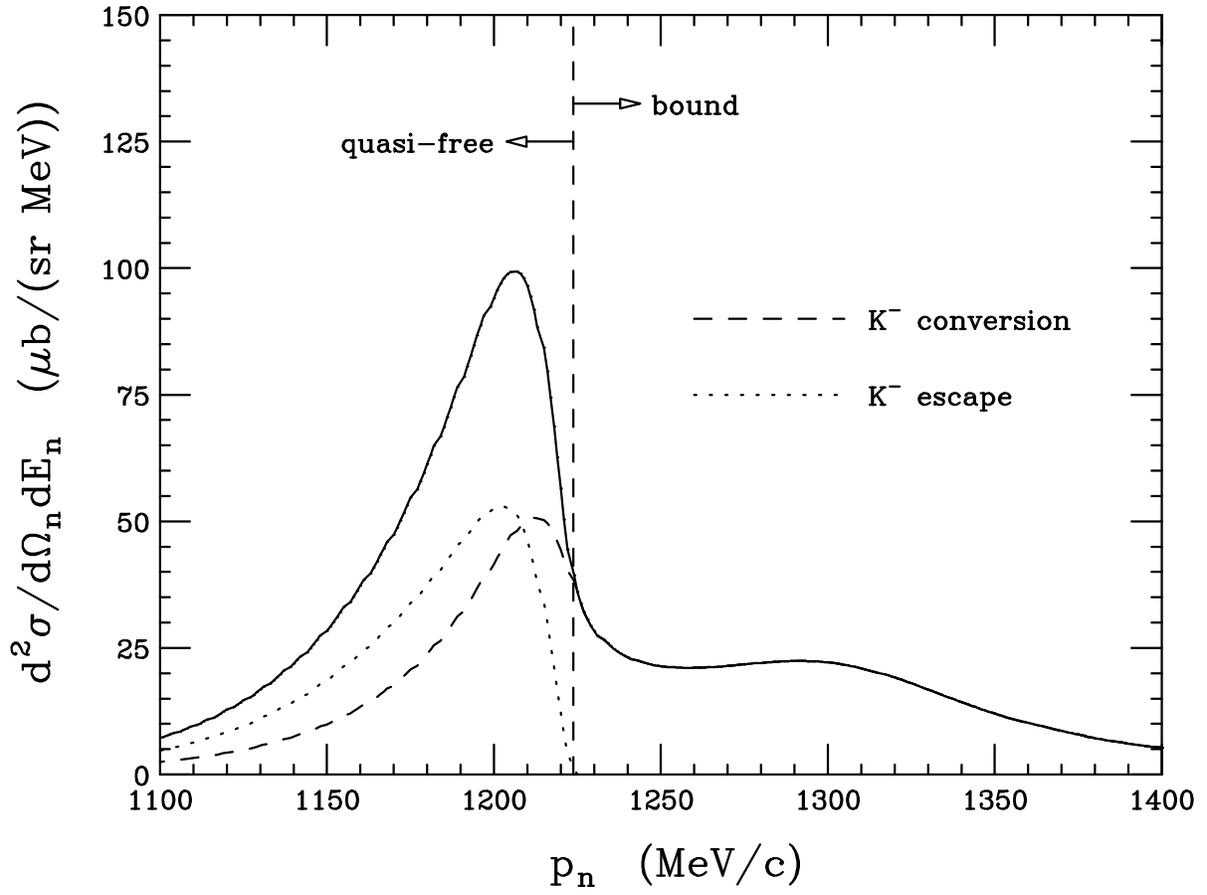}
\caption{
Same as Fig.\ref{YAspec2}, but the case of 
($V_0$, $W_0$)=($-350$ MeV, $-100$ MeV).
\label{SGMspec2}}
\end{figure}

In Fig.~\ref{YAspec2}, we display contributions of the $K^-$ 
escape and conversion components into the inclusive spectrum, 
using the YA potential.
Since the ``$pp$'' rigid-core nucleus is assumed in this calculation,
the $\bar{K}NN$ continuum spectrum is described 
as $K^-$+``$pp$'' continuum states.
Thus the $K^-$ escape spectrum in Eq.(\ref{eq:Sesc}) 
might not be fully explained in the QF region where the 
$^3$He break-up processes of 
$K^- + {^3{\rm He}} \rightarrow  n + p + p + K^-$ occur. 
On the other hand, this assumption seems to work well in the bound 
region; decay processes of 
$K^- pp \rightarrow \pi + \Sigma/\Lambda + N$ 
and $K^- pp \rightarrow \Sigma/\Lambda + N$ 
are effectively described by the imaginary potential, 
Im${U^{\rm opt}}$, if we choose an appropriate one.
Such a conversion spectrum in Eq.(\ref{eq:Scon}) 
expresses semi-exclusive 
$K^- + {^3{\rm He}} \rightarrow n + \Sigma/\Lambda + X$ 
processes detecting the selected particles in invariant-mass 
spectroscopy at the J-PARC E15 experiment.
Therefore, the treatment of the ``$pp$'' core-nucleus 
would be justified in not only the bound region but also
the continuum region, 
as long as we concern the conversion spectrum.
In Fig.~\ref{SGMspec2}, we show also the escape and conversion 
spectra using ($V_0$, $W_0$)=($-350$ MeV, $-100$ MeV), 
in order to simulate the results by Shevchenko, 
Gal and Mare$\breve{\rm s}$ \cite{She06}.
We confirm that a peak of the $K^-pp$ bound state is 
broadened because a value of $\mit{\Gamma}$ is so larger 
than that obtained by the YA potential.
The corresponding bump is barely recognized in the spectrum. 

--- Fig.2 ---

--- Fig.3 ---

\begin{figure}
\includegraphics[scale=0.7, angle=90]{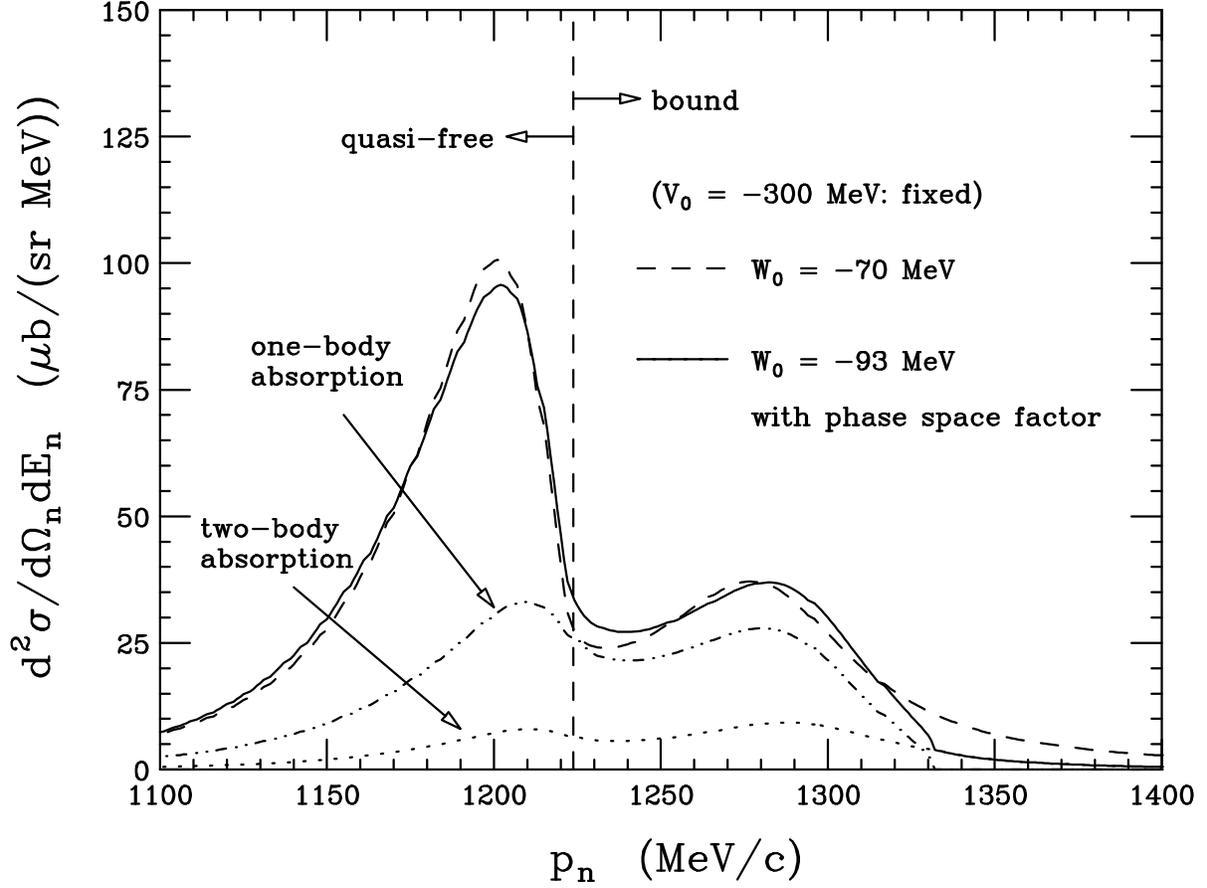}
\caption{
The effects of the phase space suppression factor in the 
$^3$He(in-flight $K^-$,$n$) reaction 
at $p_{K^-} = 1.0$ GeV/c and $\theta_n = 0^{\circ}$. 
The YA optical potential is used. 
The solid and dashed lines denote the inclusive spectra 
with and without taking into account the 
phase space factor, respectively. 
The dot-dot-dashed line indicates the contribution of 
the one-nucleon absorption process, $K^- + ^3{\rm He} \to n + \pi+Y+N$,
and the dotted line the two-nucleon one, $K^- + ^3{\rm He} \to n +Y+N$.
\label{AY-Edep}}
\end{figure}

Mare$\breve{\rm s}$, Friedman and  Gal \cite{Mar05} introduced 
a phase space suppression factor $f(E)$ multiplying
the imaginary part of the $K^-$ optical potential.
When a pole of the $K^-pp$ bound state is located below the 
$\pi \Sigma N$ threshold in the complex energy plane, 
its width would be narrower by the phase space suppression 
because the $\bar{K} N \rightarrow \pi \Sigma$ decay channel 
is closed there.
Several predictions for the $K^-pp$ bound 
state \cite{YA02,She06,Ike07,Dot07} suggest that 
its pole resides above the $\pi \Sigma N$
threshold, except the results in Ref.~\cite{Iva05}.
In order to see the effects,
we replace $W_0$  by $W_0 \times f(E)$ in Eq.(\ref{eq:KppOptPot}).
Here we assumed that the $K^-$ is absorbed from 
80 \% one-nucleon and 20 \% two-nucleons, 
as shown in Refs. \cite{Mar05,Yam06}. 
Indeed, the helium bubble chamber experiment \cite{Kat70} 
suggested that the ratio of $K^-$ absorption 
on two nucleons to all $K^-$ absorption process
in $^4$He at rest amounts to 16 \%.
Taking into account the phase space factor, we adjust 
a value of  $W_0$ to be $-93$ MeV so as to reproduce the 
same $\mit{\Gamma}$ of the $K^-pp$ bound state. 
Since $S^{\rm con}(E)$ can be further decomposed into 
the corresponding spectra from the one- and two-nucleon
absorptions,
we can estimate each component for these decay modes.
Fig.~\ref{AY-Edep} shows the effects of the phase space 
factor in the calculated spectrum. 
We find that the shape of the inclusive spectrum is not so changed 
when the phase space factor is turned on; 
the spectrum of the $K^-pp$ bound state, however,  
is sizably reduced near the $\pi \Sigma N$ threshold 
($p_n \sim 1330$ MeV/c)
because the $K^-pp$ below the threshold 
can decay into only the $K^-pp \rightarrow Y + N$ 
process from the two-nucleon absorption, 
which contributes slightly to the spectrum on the whole.

--- Fig.4 ---

\begin{figure}
\includegraphics[scale=0.7, angle=90]{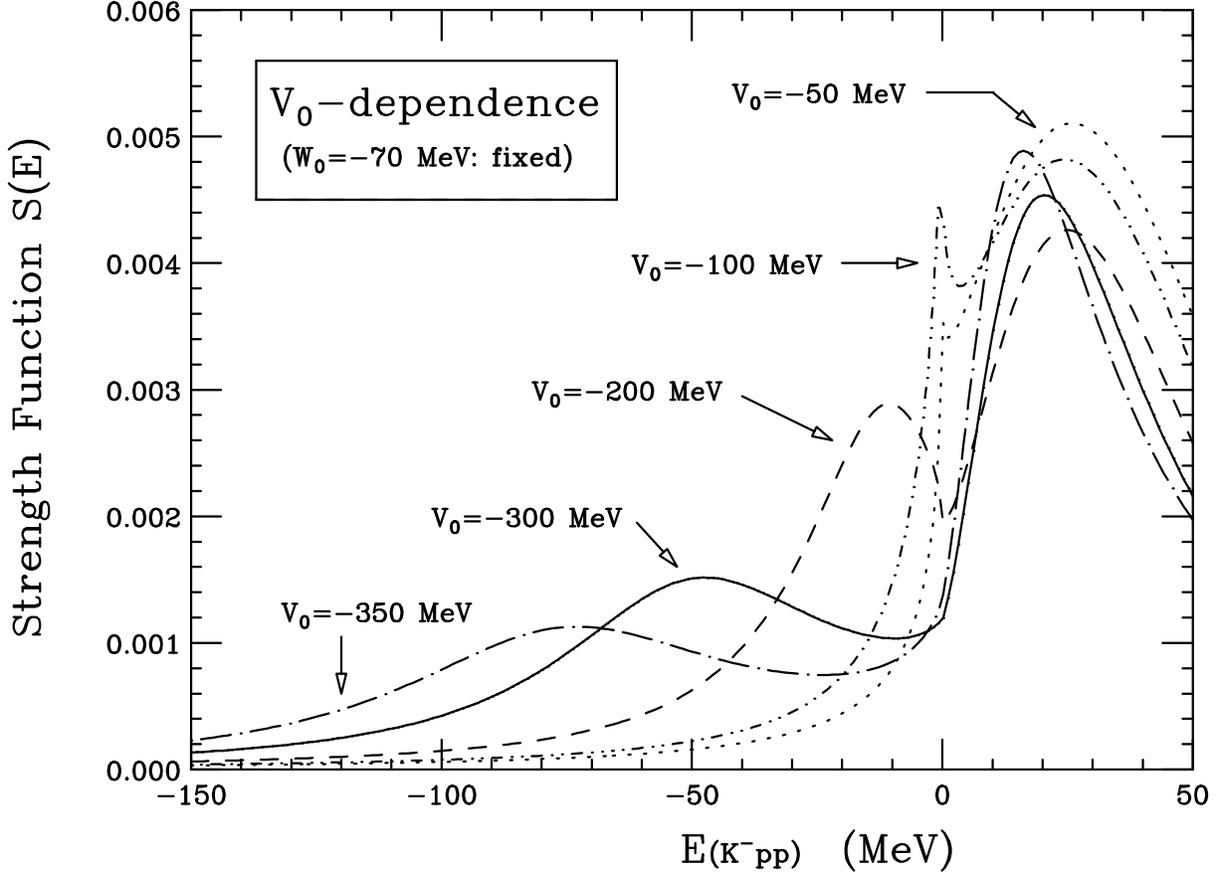}
\caption{
The behavior of the strength function 
for the $^3$He(in-flight $K^-$,$n$) reaction at $p_{K^-} = 1.0$ GeV/c 
and $\theta_n = 0^{\circ}$ 
on the strength parameter of $V_0$. 
$W_0$ is fixed to be $-70$ MeV.
The horizontal axis denotes the energy of the $K^-pp$ system 
measured from the $K^-$ emitted threshold;
$E<0$ means a bound state of the $K^-pp$ system.
\label{SVdep}}
\end{figure}

\begin{figure}
\includegraphics[scale=0.7, angle=90]{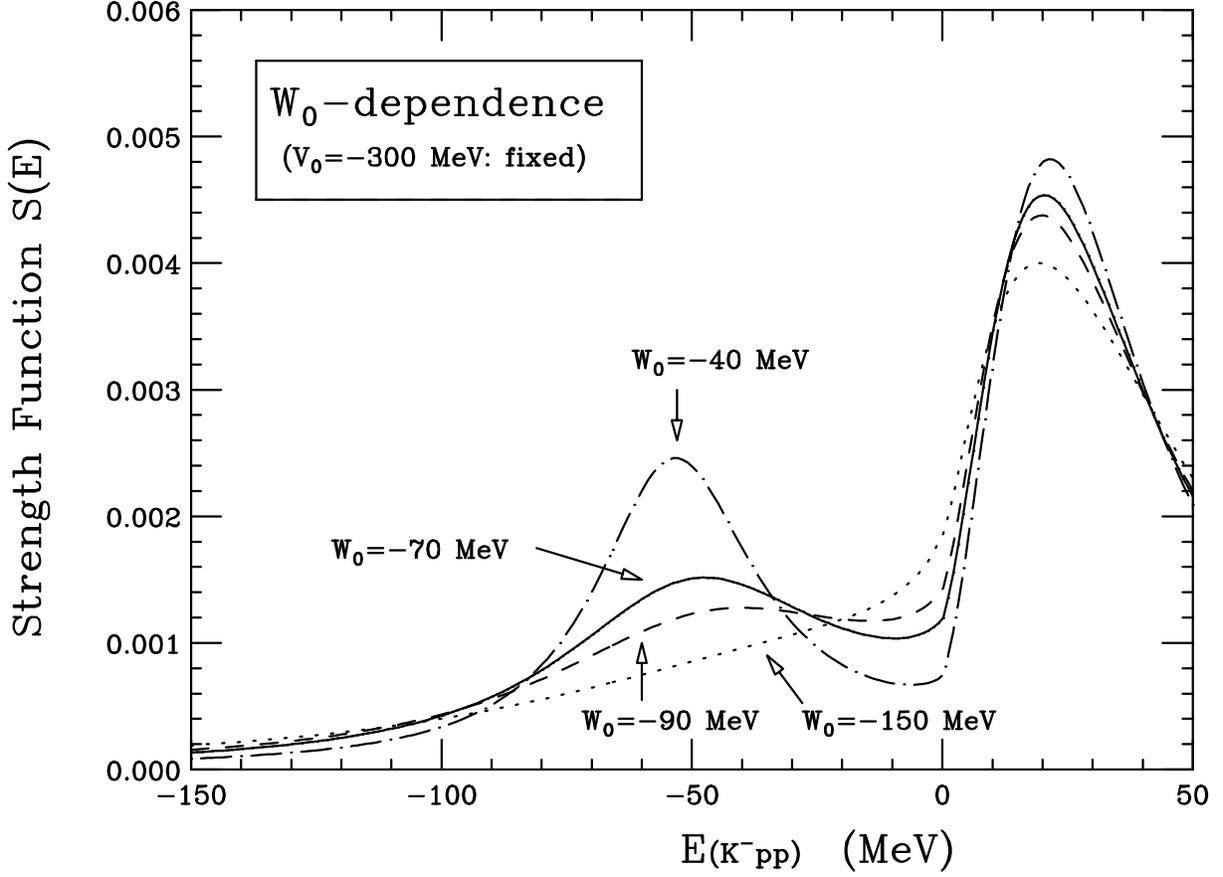}
\caption{
The behavior of the strength function 
for the $^3$He(in-flight $K^-$,$n$) reaction at $p_{K^-} = 1.0$ GeV/c 
and $\theta_n = 0^{\circ}$ 
on the strength parameter of $W_0$. 
$V_0$ is fixed to be $-300$ MeV.
See also the caption in Fig.~\ref{SVdep}.
\label{SWdep}}
\end{figure}

It is worth studying shape behaviors of the 
strength function $S(E)$ for the $K^-pp$ bound state
on the parameters of $V_0$ and $W_0$, systematically.
As shown in Fig.~\ref{SVdep}, we change the value of $V_0$ 
from $-50$ MeV to $-350$ MeV, keeping $W_0 = -70$ MeV. 
We see that a peak of the bound state gives rise to 
a cusp-like structure at $V_0=(-100)$-$(-50)$ MeV (shallow). 
In Fig.~\ref{SWdep}, we show the shape behavior of 
$S(E)$ when we change the value of $W_0$
from $-40$ MeV to $-150$ MeV, keeping $V_0 = -300$ MeV. 
We find that choosing $W_0=-150$ MeV (deep),
the peak is completely invisible even if the bound state exists.
Consequently, in order to identify a peak of the $K^-pp$ bound 
state in the inclusive spectrum, we realize that 
the value of $V_0$ must be deeper than $\sim -100$ MeV 
and also that of $W_0$ shallower 
than $\sim -100$ MeV. 
The parameter set for the YA potential \cite{YA02} 
fulfills this condition, while that simulating the results
by Shevchenko, Gal and Mare$\breve{\rm s}$ \cite{She06} 
stands on the edge of the signal observation in terms of 
the width.
We believe that it is possible to 
extract variable information on the $K^-$ optical potential from 
the shape behavior of the inclusive and semi-exclusive spectra.

--- Fig.5 ---

--- Fig.6 ---

Dot$\acute{\rm e}$ $et$ $al.$ \cite{Dot04} predicted the 
possibility of extreme nuclear shrinkage in the $K^-NNN$ system 
(strange tribaryon) by anti-symmetrized 
molecular dynamics (AMD) calculations. 
Such nuclear contraction is strongly related to the antikaon condensation 
in compact star \cite{BRL96}. 
It is interesting to evaluate the nuclear contraction 
in the $K^-pp$ bound state by the ($K^-$, $n$) reaction.
When a harmonic oscillator model is applied to $^3$He, 
the r.m.s. distance of the ``$pp$'' pair is 1.90 fm
in the $K^-pp$ bound state, whose value is reduced from 
that of 2.25 fm in $^3$He, as discussed in Ref.~\cite{YA02}. 
However, this effect is not expected to be observed because 
the cross section of the $K^-pp$ bound state is only slightly
suppressed by about 4\%.

Our simplified calculation in this paper would bring some 
uncertainties for the resultant spectrum: 
Only the $K^- + n \rightarrow n + K^-$ forward 
scattering has been considered, omitting the 
$K^- + p \rightarrow n + \bar{K}^0$ charge exchange process 
which can also contribute to the $K^-pp$ formation through 
the coupling between $K^-pp$ and $\bar{K}^0pn$ channels.
This contribution is roughly estimated 
by replacing the Fermi-averaged transition amplitude 
$f_{K^-n \to nK^-}$ for the elementary $K^- + n \to n + K^-$ 
process by the isoscalar transition amplitude
$f_{\Delta I=0} = -\sqrt{2/3} (f_{K^-n \to nK^-} + 
{1 \over 2}f_{K^- p \to n {\bar K}^0})$ 
which is equivalent to a $u$-channel isospin-0 one for 
($K^-$,$\bar{K})$ reactions, 
including a spectroscopic factor for the $K^-pp$ $I$=1/2 
state formed on $^3$He.
As a result, the cross section of the $K^-pp$ bound state is
enhanced by about 18 \% with Fermi-averaging.
It is noticed that the $K^-pp$ state with $I$=1/2 dominates 
the ($K^-$,$n$) reaction at $p_{K^-}$= 1.0 GeV/c
because $| f_{\Delta I=0}|^2/|f_{\Delta I=1}|^2 \simeq 14$, where
$f_{\Delta I=1}$ is the isovector transition 
amplitude \cite{Koike07}.
This nature justifies our assumption that 
we treat $[{\bar K}\otimes \{NN \}_{I=1}]_{I=1/2}$ as $K^-pp$ 
restrictedly in this calculation.
In order to get more quantitative results, 
a full coupled-channel calculation would be needed. 
Moreover, the choice of 
the parameters in the eikonal distorted waves also changes
the absolute value of the cross section, 
but the distortion effect is not significant for $^3$He. 
For the in-flight ($K^-$,$N$) reaction, it would be not 
appropriate to use the decay rate measured by $K^-$ absorption 
at rest,
considering that its value depends on atomic orbits where $K^-$ is 
absorbed through atomic cascade processes ~\cite{Ona89,Yam07}. 
More theoretical and experimental considerations are needed.

\section{Summary and conclusions}

We have examined the $^3$He(in-flight $K^-$,$n$) 
reaction at $p_{K^-}$ = 1.0 GeV/c and $\theta_n = 0^{\rm o}$ 
for the forthcoming J-PARC E15 experiment. 
We have demonstrated how the spectral shape of the 
inclusive spectrum is changed by the strength parameters 
of $V_0$ and $W_0$ in the $K^-$-``$pp$'' optical potential, 
and also obtained the semi-exclusive $K^-$ conversion spectrum 
which has the ability to be compared with the data 
by the experiment measuring the missing- and invariant-mass 
simultaneously. 
We believe that our considerable assumptions are adequate for 
the first step toward our purpose.
In conclusion, the necessary condition to observe a distinct peak
of the $K^-pp$ bound state turns out to be that 
$V_0$ is deeper than $\sim -100$ MeV and 
$W_0$ shallower than $\sim -100$ MeV. 
The $K^-pp$ bound state predicted by Yamazaki and Akaishi \cite{YA02} 
fulfills this condition, while that by  Shevchenko, 
Gal and Mare$\breve{\rm s}$ \cite{She06} stands on the edge of 
the signal observation in terms of the width.
To analyze in detail the forthcoming experimental data at J-PARC, 
it will be necessary to improve our theoretical treatment. 
This investigation now is in progress \cite{Koike07}.


\begin{acknowledgments}
We acknowledge Prof. M. Iwasaki for discussions about a plan 
of the J-PARC E15 
experiment and for suggestions that a theoretical calculation is 
needed for its execution.
We would like to thank Prof. Y. Akaishi, Dr. H. Ohta and Dr. H. Ohnishi 
for many valuable discussions.
This work is supported by Grant-in-Aid for Scientific Research on
Priority Areas (Nos. 17070002 and 17070007). 
\end{acknowledgments}


%
%
%
\newpage
%

%

%

\end{document}